\documentstyle[11pt,psfig]{article}
\newcommand{\s}{\rm}

\newcommand{\ra}{\rightarrow}
\newcommand{\mn}{\mu \nu}

\newcommand{\be}{\begin{equation}}
\newcommand{\ee}{\end{equation}}

\newcommand{\bea}{\begin{eqnarray}}
\newcommand{\eea}{\end{eqnarray}}
\newcommand{\bef}{\begin{figure}}
\newcommand{\eef}{\end{figure}}

\newcommand{\lgl}{\langle}
\newcommand{\rgl}{\rangle}

\begin{document}
\begin{center}
{\large{\bf {Low Mass Dileptons from Pb+Au Collisions at 158 AGeV}}}
\vskip 0.2in
{\bf Sourav Sarkar$^a$, Jan-e Alam$^{b,}$\footnote{Present Address: 
Variable Energy Cyclotron Centre, 1/AF Bidhan Nagar, Calcutta 700 064, India.} 
and T. Hatsuda$^b$}

\vskip 0.2in
\it{$^a$Variable Energy Cyclotron Centre, 1/AF Bidhannagar, 
Kolkata-700 064}\\
\it{$^b$Physics Department, University of Tokyo, Tokyo 113-0033, Japan}
\end{center}
\begin{abstract}
The medium modification of vector meson properties in hot/dense
hadronic matter and its role in explaining
the CERES/NA45 dilepton data at different centralities are discussed.
\end{abstract}

\vskip 0.2in
\addtolength{\baselineskip}{0.3\baselineskip}

Relativistic nuclear collisions are expected to produce extreme conditions
of temperature and/or baryon density conducive for the liberation of bound 
colour 
charges resulting in the formation of quark-gluon plasma (QGP). Spontaneously
broken chiral symmetry is also expected to be restored. These phenomena 
constitute a massive restructuring of the QCD vacuum and is likely to
affect the spectral properties of hadrons which are in fact excitations of
the vacuum. Electromagnetic probes can be effectively  used to study changes 
in hadronic properties in hot/dense medium because of their minimal final state
interactions. The dilepton spectra in particular, exhibits a resonant structure
which, in the low mass regime includes the $\rho$ and the $\omega$ mesons.
The spectral modifications of these vector mesons
would be clearly revealed in the invariant mass spectra of the dileptons
through the shift of the resonant peaks.

We have seen earlier~\cite{rapid} that
the WA98 photon spectra from Pb+Pb collisions at the CERN SPS
can be explained by either of the two scenarios of 
relativistic nuclear collision~\cite{red,rapid}:
(a) A\,+A\,$\ra$QGP$\ra$Mixed Phase$\ra$Hadronic Phase
or (b) A\,+\,A\,$\ra$Hadronic Matter;
with downward shift of vector meson masses
and initial temperature $\sim 200$ MeV. 
Our aim now is to check whether the same scenario can explain the dilepton
data obtained in Pb+Au collisions at SPS by the CERES/NA45 
Collaboration~\cite{ceres}.
The possible sources in the low mass region are the dileptons coming
from the decays of hadrons at freeze-out and from the in-medium
propagation and decay of vector mesons. The data shows a significant
enhancement in the mass region 0.3 to 0.6 GeV which can 
be explained~\cite{lowmass} by a substantial negative shift of the mass 
of vector mesons (the $\rho$
meson in particular) in the thermal medium.
A large broadening of the $\rho$ meson spectral function due to scattering off
baryons can also explain this enhancement~\cite{rapp}
though the photon yield is seen to be insensitive to such a 
broadening~\cite{annals}.

A substantial volume of literature has been devoted  to the
issue of temperature and/or density dependence of hadrons within
various models~\cite{thpr,brpr,rapp,rdp}. 
The effects of the thermal shift of the hadronic
spectral functions on both photon and dilepton emission have
been considered in Ref.~\cite{annals}
for an exhaustive set of models.  An appreciable
change in the space-time integrated yield of electromagnetic
probes was observed for universal scaling~\cite{brpr}
and Quantum Hadrodynamic (QHD) model~\cite{vol16}
 and we will consider only these in the present discussion.

Let us now briefly recapitulate the main
equations relevant for evaluating dilepton emission
from a thermal source.
The emission rate of dileptons 
can be expressed in terms of the imaginary part of
the retarded electromagnetic current correlation function
as~\cite{emprobe}
\be
\frac{dR}{d^4p}=-\frac{\alpha}{12\pi^4\,p^2}
\,g^{\mn}\,{\s Im}W^R_{\mn}(p)
\left(1+\frac{2m_l^2}{p^2}\right)\,\sqrt{1-\frac{4m_l^2}{p^2}}
\,f_{BE}(E,T)
\label{drlep}
\ee
where $m_l$ is the lepton mass.

\bef
\centerline{\psfig{figure=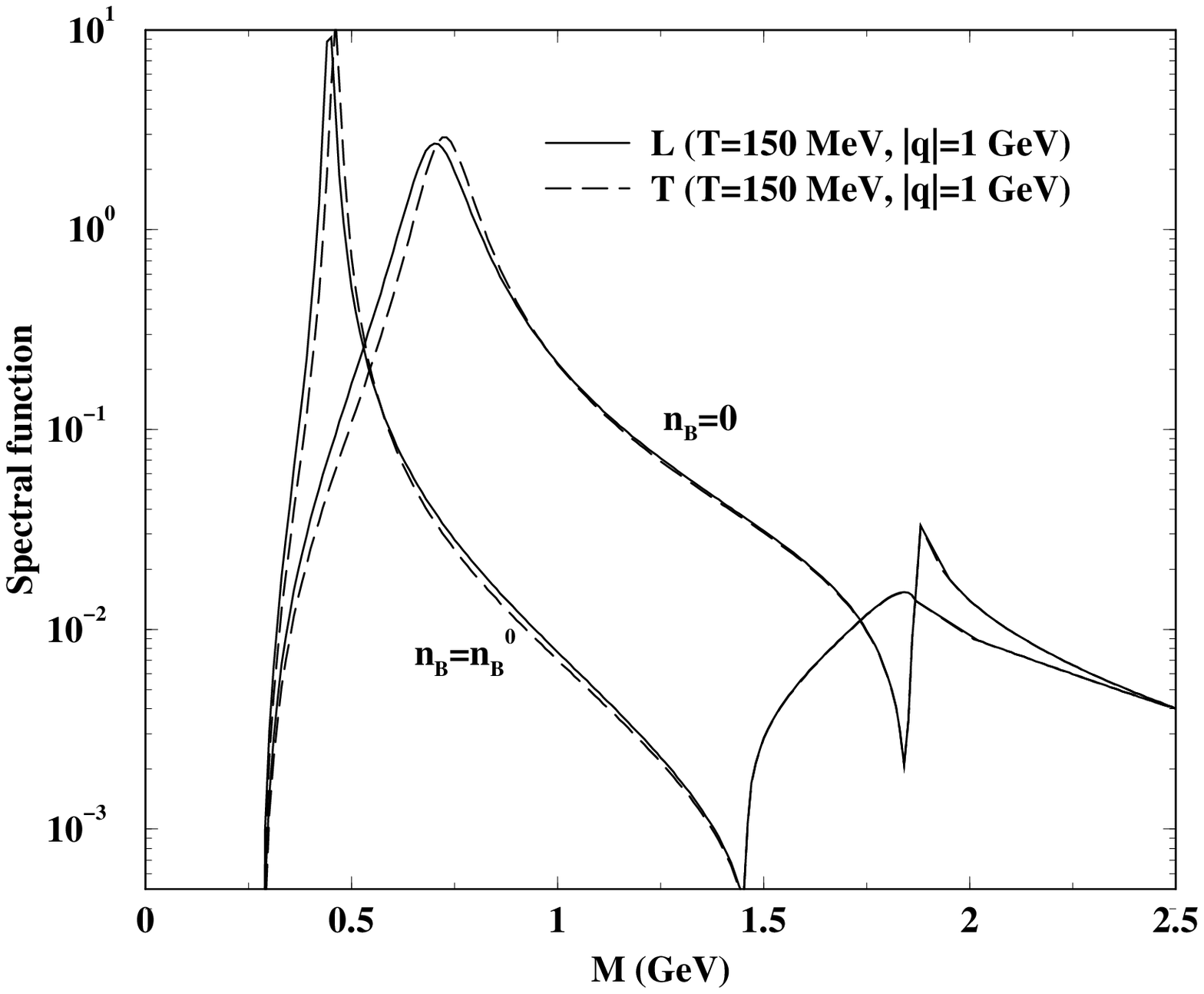,height=7cm,width=8cm}}
\caption{
The spectral function of the $\rho$ meson in Quantum Hadrodynamics (QHD).
}
\label{spec}
\eef

In the QGP, the dominant channel for dilepton productions is
quark-antiquark annihilation.
The rate for this process is obtained from the lowest order
diagram contributing to the current correlator $W^{\mn}$.
In order to obtain the rate of dilepton emission from hadronic matter
($\rho/\omega\ra l^+l^-$) from Eq.~(\ref{drlep})
the electromagnetic current correlator is expressed in terms of
the effective propagator of the vector particle in the thermal medium
using vector meson dominance (VMD) so as to obtain
\bea
{\s Im}W_{\mu\mu}^R&=&\frac{\pi e^2 m_V^4}{g_V^2}(2\rho^V_T+\rho^V_L)
~~;~~~~~{\s with}\nonumber\\
\rho^V_{T(L)}&=& \frac{1}{\pi}\left[\frac{{\s Im}\Pi^V_{T(L)}}
{(p^2-m_V^2+{\s Re}\Pi^V_{T(L)})^2 +({\s Im}\Pi^V_{T(L)})^2}\right]
\eea
where $\rho^V_{T(L)}$ and $\Pi^V_{T(L)}$ 
are respectively the transverse (longitudinal) parts of
the retarded spectral function and self-energy of vector mesons 
arising out of interaction with excitations in the medium. 
The transverse and longitudinal components of the
spectral function of the $\rho$ with interactions from the QHD model is shown
in Fig.~\ref{spec}. The downward shift of the
first peak of the spectral function with increasing
density $n_B$ and temperature $T$ can be almost entirely
attributed to the $\bar N N$ polarisation of $\rho$ in the modified Dirac sea.
The narrowing down of the width with increasing $T$ and $n_B$
is due to the reduction of the phase space for the
process, $\rho\rightarrow \pi\pi$
resulting from the decrease in the $\rho$ mass.
The second peak in the spectral function is due to the
$\bar{N}N$ threshold of the off-shell $\rho$ which has negligible
effects on the dilepton yield. For a discussion on the $\rho$ spectral
function in the scaling scenario see Ref.~\cite{annals}.

The observed dilepton spectrum originating from an expanding
matter can be obtained by convoluting the static
rates given above with the expansion dynamics. 
This is done using boost invariant relativistic hydrodynamics.
For the QGP sector we use the bag model equation of state with
two flavour degrees of freedom. 
The hadronic phase is taken to consist of $\pi$, $\rho$, $\omega$,
$\eta$ and $a_1$
mesons and nucleons.
The medium effects enter through the effective masses in
the expressions for energy density and pressure.
The entropy density is then parametrized as,
\be
s_H=\frac{\epsilon_H+P_H}{T}\,\equiv\,4a_{\s{eff}}(T)\,T^3
= 4\frac{\pi^2}{90} g_{\s{eff}}(m^\ast,T)T^3
\label{entro}
\ee
where  $g_{\s{eff}}$ is the effective statistical degeneracy through
which medium effects enter into the expansion dynamics of the system.
The velocity of sound consequently
becomes a function of $T$ and differs substantially from its value
corresponding to an ideal gas ($1/\sqrt 3$). 
When the system is produced in the QGP phase, $g_{\s{eff}}$ 
is replaced by $g_{QGP}$ which
for two quark flavours is 37. 
Note that the value of the initial temperature 
of a hot hadronic gas also depends on $g_{\s{eff}}$.
The freeze-out temperature $T_F$ is taken as 120 MeV.

\bef
\centerline{\psfig{figure=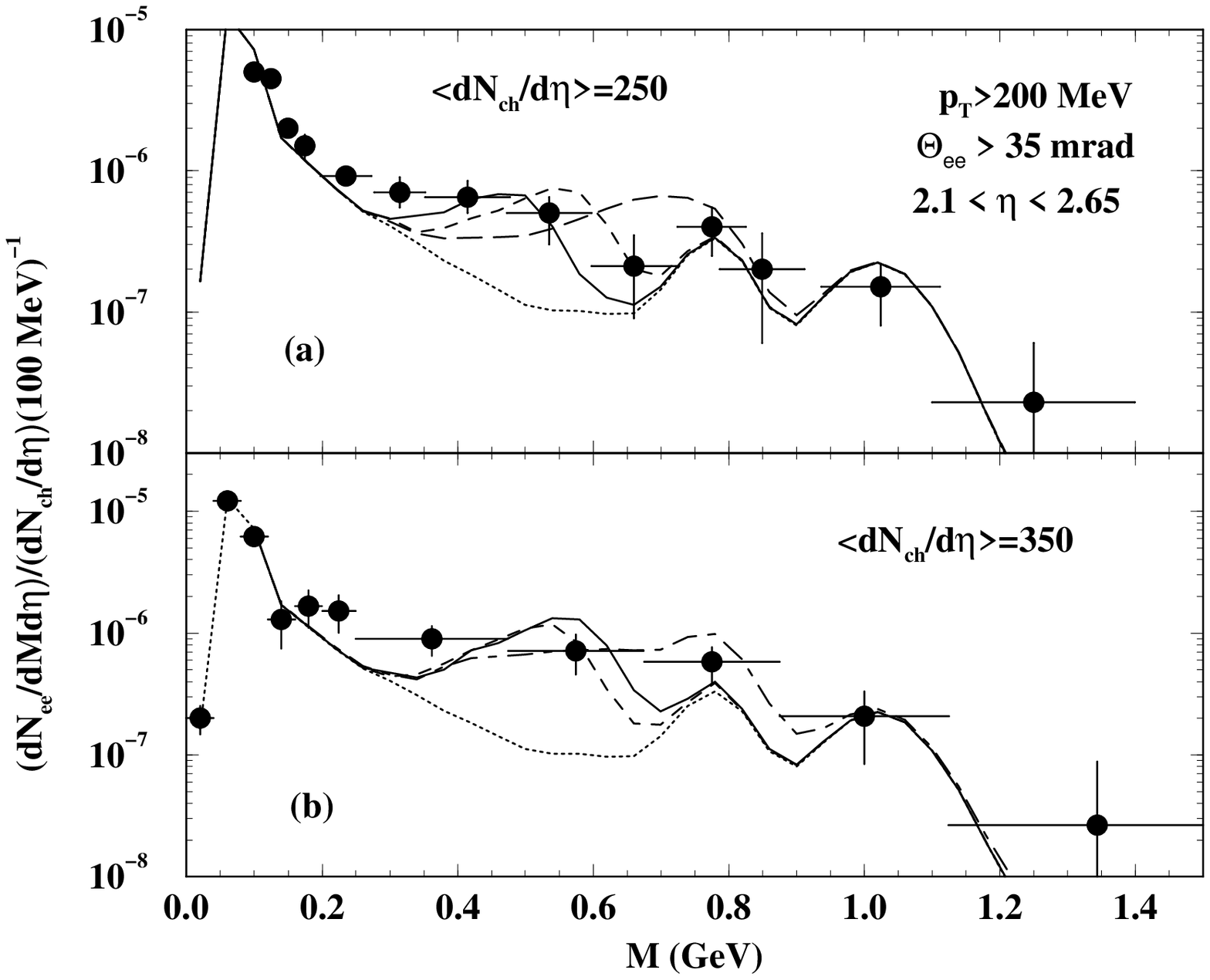,height=7cm,width=8cm}}
\caption{Dilepton spectra with
different initial states and mass variation scenarios 
for (a) $\lgl dN_{ch}/d\eta\rgl$=250 and (b) $\lgl dN_{ch}/d\eta\rgl$=350.
}
\label{fig_ceres}
\eef
We now compare the results of our calculation with the dilepton 
spectra obtained by the CERES/NA45 Collaboration in Fig.~\ref{fig_ceres}.
The upper panel (a) contains the results for $\lgl dN_{ch}/d\eta\rgl$=250
where we have considered a hadronic gas
scenario as well as a deconfined initial state with $T_i=180$ MeV
which evolves into a hadronic gas via a mixed phase.
The observed enhancement of the dilepton yield
around $M\sim 0.3 - 0.6 $ GeV can be reproduced
with a QGP initial state once the variation
of vector meson masses in the mixed and the hadronic phases
are taken into account (solid line).
The data is also reproduced
by a hadronic initial state with $T_i=190$ MeV 
in the universal mass variation scenario (dashed line).
The shift of the $\rho$-peak in the dilepton spectra towards
lower invariant mass $M$ is more for the universal scaling scenario
compared to QHD model (long dashed line)
indicating a stronger medium effect in the former case. 
The dilepton spectra for $dN_{ch}/d\eta=350$ is shown
in the lower panel (b) of Fig.~\ref{fig_ceres}.
Calculations with an
hadronic initial state and universal scaling of hadronic masses
with temperature (dashed line) describes the data reasonably well.
The initial temperature in this case is $\sim$ 195 MeV.
We have seen that with the temperature dependent mass from QHD
model the low mass enhancement of the experimental
yield cannot be reproduced (curve not shown).
A good description of the data
can also be obtained by taking $T_i=200$ MeV
with QGP initial state for $T_c=190$ MeV (solid line).
For comparison, we show the results due to a large broadening
of the $\rho$ spectral function in the medium (dash-dotted line). 
The broadening
of $\rho$ is modelled by assuming the temperature dependent
width as : $\Gamma_\rho(T)=\Gamma_\rho(0)/(1-T^2/T_c^2)$.
In all the cases background due to hadron
decays (dotted line) are added to the thermal yields.

In summary, we have evaluated the low mass dilepton yield
in Pb + Au collisions at
CERN SPS energies with different initial conditions
for two values of the charge multiplicity.
The effects of the variation of hadronic masses
on the dilepton yield have been considered both in the cross section
as well as in the equation of state.
It is
observed that 
the data can be described by both QGP and hadronic
initial states with an initial temperature $\sim$ 200 MeV
which is very much in agreement with our conclusion from the
analysis of the WA98 photon data~\cite{rapid}. 
We have assumed $\tau_i=1$ fm/c at SPS energies,
which may be considered as the
lower limit of this quantity because
the time taken by the nuclei to pass
through each other in the CM system is $\sim$ 1 fm/c at SPS
energies and the thermal system can be formed after this time
has elapsed.

{\bf Acknowledgement:} J.A. is grateful to the JSPS and the University of Tokyo for their hospitality.

\end{document}